\documentclass[showpacs, aps, nofootinbib]{revtex4}
\usepackage{amsmath,amssymb,amsthm}
\usepackage{graphicx}
\usepackage{latexsym}

\def\demi{\frac{1}{2} }
\def\C{\mathbb{C}}
\def\P{\mathbb{P}}

\newcommand{\R}{\mathbb{R}}

\def\ff{{\cal F}}

\def\nn{{\cal N}}

\def\ppp{{\cal P}}

\def\vv{{\cal V}}

\newcommand{\be}{\begin{equation}}
\newcommand{\ee}{\end{equation}}
\newcommand{\bes}{\begin{eqnarray}}
\newcommand{\ees}{\end{eqnarray}}
\def\cop{\Delta}
\newcommand{\ket}[1]{|#1\rangle}

\def\arr{\rightarrow}

\newcommand{\lalg}[1]{\mathfrak{#1}}
\newcommand{\SU}{\mathrm{SU}}
\newcommand{\U}{\mathrm{U}}
\newcommand{\SO}{\mathrm{SO}}
\newcommand{\ISO}{\mathrm{ISO}}
\newcommand{\mat}[2]{\left(\begin{array}{#1} #2\end{array}\right)}
\newcommand{\Ref}[1]{(\ref{#1})}
\def\arr{\rightarrow}
\def\f{\frac}
\def\v{\overrightarrow}
\def\la{\langle}
\def\ra{\rangle}
\def\tl{\widetilde}
\def\nn{\nonumber}

\begin{document}
\title{Special Relativity as a non commutative geometry:\\ Lessons for Deformed Special Relativity }
\author{{\bf Florian Girelli}\footnote{fgirelli@perimeterinstitute.ca}, {\bf Etera R. Livine}\footnote{elivine@perimeterinstitute.ca}}
\affiliation{Perimeter Institute, 35 King Street North Waterloo, Ontario Canada N2J 2W9}

\begin{abstract}
\begin{center}
{\small ABSTRACT}
\end{center}
Deformed Special Relativity (DSR) is obtained by imposing a maximal energy to Special Relativity and deforming the Lorentz symmetry (more exactly the Poincar\'e symmetry) to accommodate this requirement. One can apply the same procedure deforming the Galilean symmetry in order to impose a maximal speed (the speed of light). This leads to a non-commutative space structure, to the expected deformations of composition of speed and conservation of energy-momentum. In doing so, one runs into most of the ambiguities that one stumbles onto in the DSR context.
However, this time, Special Relativity is there to tell us what is the underlying physics, in such a way that we can  understand and interpret these ambiguities. We use these insights to comment on the physics of DSR.
\end{abstract}

\maketitle

\tableofcontents

\section*{Introduction}

Deformed Special Relativity (DSR) \cite{DSR} has been proposed as a generalization of Special Relativity.
Instead of having only a maximal  speed, one also imposes a maximal energy\footnote{More precisely, the DSR theory takes into account a universal energy scale, which implies a bound of energy or mass or momentum according to the particular version of the theory.}. In order to respect such a maximal energy, one has to deform the symmetries, therefore obtaining DSR. The space-time underlying this theory (constructed as the dual of the momentum space) can be seen as a non commutative geometry or, alternatively, as carrying a metric observer dependent. The two constructions are {\it a priori} physically inequivalent and it is still unclear which one should be preferred. The ambiguity essentially lies in the reconstruction of space-time from the momentum space structure.  There are also other ambiguities like the right physical definition of energy and the notion of speed,  and issues like how one should recover the classical framework for macroscopic objects with the usual definition of energy-momentum. It is definitely important to understand and to cure these ambiguities as DSR is supposed to describe some effective limit of quantum gravity as many arguments hint to \cite{dsrcosmo, dsr3d, qg-effect}.

As we said, DSR arose when one uses the algebraic tools of non-commutative geometry in order to introduce a universal maximal energy in Special Relativity. This latter contains already a maximal speed, so a natural question one may ask is: "Is it possible to start with a Galilean space and then introduce a maximal speed by deforming the Galilean symmetries?" More strikingly one could wonder what would have happened if Einstein didn't have the Lorentz symmetries at hand and didn't think about unifying space and time to construct Special Relativity, would he have constructed some kind of non-commutative deformation of galilean symmetries in order to deal with the assumption of a universal conserved speed?

We propose here to describe Special Relativity as a non commutative geometry of space. We then take this as a guiding principle to deal with the problems arising in DSR. Namely this alternative description of Special Relativity can be constructed in a similar way as DSR and, therefore, runs into the same ambiguities. However in this case we have  Special Relativity to guide us to solve those ambiguities and provide us with a consistent physical interpretation of the mathematical construction.

This points to  a general  scheme for building new physical theories: first one determines the symmetries
that are inherent to the studied physics, then later one discovers that physics have some quantities which should be
bounded by a universal constant (i.e. the same for all observers) at some scale. This is generically not compatible with the original symmetries. Then one has a choice: either be "conservative" and (quantum) deform the symmetries in such a way to impose maximal values to these quantities, or be "innovative" and introduce some new kind of symmetries, in some sense bigger as they must encompass the previous symmetry.

In the early XX century, people have followed the latter path to build Special Relativity.  Deformed Special Relativity was created following the first. Here, we try to stay "conservative" and consider the (non-commutative) deformation approach to deal with the physics of  Special Relativity.

Note that our construction is essentially different than the one in \cite{deformedgalilee}, where the
deformation of the Galilean group is obtained as a limit from the deformation of the Poincar\'e group.

\medskip

In the next section, we first construct the deformed space of speeds constructed to account for a modified law of speed composition with a invariant maximal speed.
Section II deals with the physics of Special Relativity as seen from this non-commutative space point of view. We describe the non-commutative geometry of space and comment on the deformed associativity of the theory. We introduce the new definitions of energy and momentum, which are consistent with the deformed space of speeds. The ambiguities of the theory lay in its physical interpretation: which are the physical quantities one measures in experiments? We address these ambiguities using our knowledge of Special Relativity. We finally tackle particle scattering and describe energy-momentum conservation in our introduced non-commutative space.
Section III introduces the concept of a relative non-commutative geometry, which provides a conceptual and mathematical framework to deal with the fact the description of space(time) and events will depend on the observer and its choice of coordinates (i.e the physical quantities he does measure): the algebraic structure of the non-commutative geometry will depend on the observer.
Section IV applies the insights obtained through the previous analysis of Special Relativity. We draw some lessons for the DSR case. We comment on how to fix the ambiguities in the definition of physical notions in DSR. We comment on the modified associativity law resulting from our point of view and introduce a Lorentz precession in DSR as an analog of the Thomas precession of SR. We comment on the five-dimensional symmetries of the 4d non-commutative spacetime.
We finally conclude with speculations on the potential link between DSR and Born-Infeld kinematics\cite{BI}, which proposes the alternative way to take into account a maximal energy by enlarging the symmetry group instead of deforming it.

\section{From Galilean space to a non-commutative space }
\label{section1}

In this section, we present {\it Special Relativity} as a deformation of the Galilean framework. While Special Relativity introduces the concept of a unified spacetime, {\it Time} and {\it Space} are distinct entities in the the Galilean or Newtonian approach. Therefore we propose to look at Special Relativity and reformulate it from a three-dimensional point of view: speeds are still defined as the derivatives of the space positions with respect to the absolute time, and momenta are simply proportional to them. The original symmetry is simply $\ISO(3)$.  In this context, the speed addition, when composing reference frames,  is given by the usual simple relation:
\begin{equation}
\v{v_1}+\v{v_2} = \v{v}_{tot}.
\end{equation}
This law makes clear that there is no maximal speed. We now want to introduce one, as a matter of fact the speed of light, noted as usual $c$. To implement the maximal speed, we have  the choice of proceeding - two methods which are actually equivalent:  we can either start by looking for the most general deformation of the co-product of the $\ISO(3)$ group, just as done in {\it Deformed Special Relativity} (DSR) and then derive the whole class of deformations consistent with the existence of a maximal speed, or we can use the geometrical picture of the homogenous space and stipulate that speeds now live on the curved 3-hyperboloid and not in flat space anymore.
We chose the latter, which is rather natural from the traditional presentation of Special Relativity.
We shall check in the next section that this new structure for the speed space respects the maximal speed.

In the Galilean framework, the speed space is identified as the quotient space $\ISO(3)/\SO(3)\sim \R ^3$, as the speeds can be considered as the generators of translations in the three-dimensional space.
In order to have a maximal speed, we introduce some curvature  and  deform $\R^3$ into the hyperboloid $\SO(3,1)/\SO(3)$. As an argument for this choice, one can advocate the fact that we know from Special Relativity that the momentum space is a hyperboloid (on the mass-shell), so the hyperboloid structure seems a natural choice\footnotemark.

\footnotetext{We do not seek to rediscover the theory of Special Relativity, but merely to reformulate it algebraically from the three-dimensional point of view, showing how its mathematical structure is close to Doubly/Deformed Special Relativity (DSR). Then, instead of $\SO(3,1)/\SO(3)$, one could also try using  the 3-sphere $\SO(4)/\SO(3)$. However the construction then does not lead to the right structure: one would derive a discrete space and some unnatural commutations relations between momenta and position. From this perspective, the non-compact group $\SO(3,1)$ has to be preferred to $\SO(4)$.}

\medskip

Our construction follows the one introduced by Snyder \cite{snyder} in the DSR context. Let us
consider the homogenous space $\SO(3,1)/\SO(3)$, which we want to interpret as the space of speeds.
The 3d hyperboloid is defined through its embedding in the 4d Minkowski space:
\begin{equation}
1= \alpha_0 ^2- \alpha_1 ^2- \alpha_2 ^2-\alpha_3 ^2.
\label{hyperbol}
\end{equation}
The group $\SO(3,1)$ acts naturally on this space
\begin{equation}\label{actionso4}
\begin{array}{rcl}
[L_i, \alpha_j]&=& i\epsilon_{ijk}\alpha_k,  \; [L_i, \alpha_0]=[L_i, \alpha_3]=0, \\
{[}N_i, \alpha_j{]}&=& \delta_{ij}\alpha_0,  \; [N_i, \eta_0]= i\alpha_i,  \; [N_i, \alpha_3]=0,
\end{array}
\end{equation}
where $L_i=\epsilon_{ijk}L_{ij}$ and $N_i=L_{i0}$, $L_{ij}\in SO(3,1)$ respectively generates the (space) rotation and the boosts. The hyperboloid is identified with the space of speeds, which is isomorphic to the space of (galilean) momenta (up to a scale factor). The speed is defined through a choice of a coordinate system, which we express in terms of the Minkowski variables $\alpha_i$. The most general (isotropic) expression is given by
\begin{equation}
v_i \,=\, c\frac{\alpha_i}{\alpha_0}F(\alpha_0).
\end{equation}
The coordinate system is the link which translates the mathematical objects $\alpha_\mu$ into the physical object defined as the speed $v$ (and measurable experimentally). Then, injecting these coordinate systems in the algebra (\ref{actionso4}), $F$ labels the deformation at the algebraic level. Some coordinates might be preferred in the sense that the (Hopf) algebra takes a particular form or that the coordinates acquire an appropriate physical meaning. A natural choice is  $F=1$ which gives the {\it Snyder coordinates}. This choice is also the traditional definition of
the speed in Special Relativity: $\alpha_\mu$ is the relativistic speed 4-vector, $\alpha_0$ defines the relativistic factor $\gamma$ while the space speed 3-vector is exactly given by $v_i=c\,(\alpha_i/\alpha_0)$. In this case, \Ref{hyperbol} gives the usual relation:
$$
\gamma\equiv \alpha_0=\f{1}{\sqrt{1-\f{v^2}{c^2}}}
$$

Let us introduce at this point some further useful notations. To parametrize the hyperboloid, one commonly introduces the boost parameter $\eta$ defined by $\alpha_0=\cosh\eta$ and the normalized boost direction $b_i=\alpha_i/\sinh\eta$, so that the (Snyder) 3-speed is expressed as $\v{v}=\,c\,\tanh\eta\,\v{b}$.
More generically, for an arbitrary deformation function $F$, we have $\v{v}=\,c\,F(\eta)\tanh\eta\,\v{b}$, and the speed is bounded by definition as long as $F$ is bounded. From this perspective, a physically reasonable choice of $F\times\tanh$ is a bounded increasing function.


\medskip

We reconstruct the 3d (coordinate) space as the tangent space to the speed space. More precisely, the (Cartesian) coordinates will be the (Hermitian) generators of the translation on the speed space (up to a dimensional scale factor), which are actually the boost operators:
\begin{equation}
x_i \,\equiv\, \f{\hbar}{mc} N_i \,=\, i\f{\hbar}{mc}  \left(\alpha_0\frac{\partial}{\partial \alpha_i}+\alpha_i\frac{\partial}{\partial \alpha_0}\right).
\end{equation}
We have introduced the (reduced) Planck constant $\hbar$ and the (rest) mass $m$ of the considered system/particle. The pre-factor $\hbar/mc$ is actually the Compton length $l_C$ associated to the system and is here for the purpose of dimensional analysis.

The rotation transformations act normally on these coordinates and are naturally represented as:
\be
L_{ij}\,=\, i\left(\alpha_i\f{\partial}{\partial \alpha_j}- \alpha_j\f{\partial}{\partial \alpha_i}\right)\\
\ee
The new feature is that, even though the coordinates operators transform indeed covariantly under space rotations, the space coordinates are now non-commutative and their bracket is given by
\begin{equation}
[x_i, x_j]\,=\, i  \left(\f{\hbar}{mc}\right)^2 L_{ij} \,=\, i  l_C^2 L_{ij}.
\end{equation}
Note that the position operators have a continuous spectrum. Then it is obvious that when $c\rightarrow\infty$ we
recover the usual three-dimensional commutative space. We have constructed here the analogue of the $\kappa$-Minkowski space, which we name the $c$-Galilean space.
Let us comment on the uncertainty relations resulting from the non-commutativity of the space coordinates. To localize a point in this 3-space, we need to build coherent states which will be considered as semi-classical points \cite{discrete}. A measure of the uncertainty in the coordinates can be defined as $(\delta x)^2=\la x_ix_i\ra-\la x_i\ra\la x_i\ra$. Taking into account that the Hilbert space of the theory is the space of functions over the hyperboloid, which can be decomposed over the simple representations of the Lorentz group $\SO(3,1)$. Such representations are of two types: a discrete series labeled by an integer $n$, which can be decomposed into the $\SO(3)$ representations of spin $j\ge n$, and a continuous series labeled by a real positive parameter $\rho$, which decomposes into all spins $j\ge 0$. Computing the uncertainty of states of fixed spin $j$ in a given representation $n$ or $\rho$, it is straightforward to check that $\la j,m| x_i|j,m\ra=0$ and that $(\delta x)^2$ is simply the difference of the Casimir operators of $\SO(3,1)$ and of $\SO(3)$:
$$
(\delta x)^2=\,l_C^2\times (j(j+1)+\rho^2+1)
\quad\textrm{or}\quad
l_C^2\times (j(j+1)-n^2+1).
$$
It is easy to see that the minimal indeterminacy is reached for $j=n$ when $\delta x=l_C\,\sqrt{n}$. This way, we have derived that one can not localize the particle/system with a better resolution than its Compton length, which is the expected physical effect when introducing the Planck constant $\hbar$ in the realm of Special Relativity.

\medskip

One way to understand the non-commutativity of the space coordinates is to realize that we have traded the loss of absolute simultaneity arising in Special Relativity against the loss of exact localization in our framework. Indeed, in Special Relativity, the notion of simultaneity depends on the observer, whose movement defines a specific foliation of spacetime into (3d) space slices. Changing the speed of the observer will affect the choice of foliation and will therefore modify the coordinates the observer assigns to objects. In our framework, we have assumed a universal time and a specific choice of splitting of spacetime into a unique notion of 3d space and a fixed time direction. Starting from Special Relativity, the choice of observer modifies measurements of coordinates but we choose to ignore this so that the indeterminacy of which space slicing we choose gets translated into a fuzziness of the space coordinates. Let us make this argument more precise and consider a (static) object in the 3d space whose coordinates we want to measure. Determining the coordinates is not a straightforward measurement like measuring the distance to the object would be. Indeed measuring the distance simply involves sending a ray of light to the object and the time of the return flight, where as measuring the coordinates actually involves a motion on the behalf of the observer. Starting from the origin, wanting to calculate the coordinate $x$ of the object, one will move along the $x$ axis, constantly measure the distance to the object so that $x$ will be determined as the point $X$ on the axis with minimal distance. Then starting from $X$, one can move along the $y$ axis in order to determine the coordinate $y$. As the observer is in motion, in the context of special relativity, the measurements will involve a change of space slicing, basically a boost in the $x$ direction followed by a boost in the $y$ direction. From this point of view, if we had measured $y$ first and then only $x$, we would have done a boost in the $y$ direction first and the a boost in the $x$ direction. As boosts do not commute, we would not end up with the same space slicing at the end of the two measurements, so that the actual measured values of $x$ and $y$ would be different in the two experiments: $x$ and $y$ do not commute if we decide to stipulate a universal splitting of spacetime into fixed space and fixed time and ignore the dependence of the space slicing on the motion of the observer.

\medskip

We can compute the brackets of the speed/momentum with the space coordinates straightforwardly once the deformation function $F$ is specified. Explicitly, in the case that $F=1$, we get:
\begin{equation}
[x_i,v_j]\,=\,i\f{\hbar}{m}\left(\delta_{ij}-\frac{\alpha_i\alpha_j}{\alpha_0^2}\right)
\,=\, i\f{\hbar}{m}\left(\delta_{ij}-\frac{v_iv_j}{c^2}\right).
\label{poisson}
\end{equation}
When $|v|\ll c$, then this reduces to the standard phase space bracket $\{x_i,p_j=mv_j\}=\hbar\delta_{ij}$. However, when $|v|\arr c$, then this goes to $\{x_i,p_j\}=0$. Let us also point out that we have off-diagonal terms: $x_1$ doesn't commute with $v_2$. If we now compute the commutators with the relativistic 4-speed, we get the somewhat familiar relations:
$$
[x_i,\alpha_0 mc^2]=\,i\,\hbar c\,\alpha_i=\,i\,\hbar\,\alpha_0 v_i,
\qquad
[x_i,mc\alpha_j]=\,i\,\hbar\,\alpha_0\delta_{ij}.
$$
As we will discuss later, $(\alpha_0 mc^2,\alpha_i mc)$ defines a new notion of energy-momentum needed to accommodate the principle of relativity in this deformed Galilean framework, and actually matches the usual relativistic energy-momentum 4-vector.

\medskip

Let us illustrate these previous considerations in the simplified two-dimensional case. As we have only one spatial coordinate, one can not expect a non-commutative space. The hyperboloid is now one-dimensional and is defined as the homogenous space $\SO(1,1)$. Its embedding in the 2d Minkowski space is given by
\begin{equation}
1= \alpha_0 ^2- \alpha_1^2.
\end{equation}
We then define:
\begin{equation}
\left|
\begin{array}{rcl}
v &=& c\,\frac{\alpha_1}{\alpha_0}F(\alpha_0),\\
x &=& i  l_C\,\left(\alpha_0\frac{\partial}{\partial \alpha_1}+\alpha_1\frac{\partial}{\partial \alpha_0}\right).
\end{array}
\right.
\end{equation}
The Snyder basis is defined as the $F\equiv 1$ choice, for which we have $v=c(\alpha_1/\alpha_0)$. The speed can then be more simply parametrized as $v=c\,\tanh \eta$ using the hyperbolic angle or boost parameter $\eta$ defined through
$\alpha_0=\cosh\eta, \alpha_1=\sinh\eta$. The bracket between the coordinate and the speed takes the simple form:
\begin{equation}\label{px2d}
[x, v]=i\,\f{\hbar}{m}\left(1-\frac{v^2}{c^2}\right)
\quad\textrm{or equivalently}\quad
[x,p]=\, i\,\hbar\left(1-\frac{v^2}{c^2}\right).
\end{equation}
It is clear that we recover the simple bracket between $x$ and the momentum $p=mv$ only when $|v|\ll c$ while it goes to $[x,p]\arr 0$ when $|v|\arr c$. Moreover, in this simplified model, it is straightforward to extract a "true" conjugate variable to the coordinate $x$. Indeed, we have:
\be
[x,\eta]=\, i \f{\hbar}{mc}=\,i\,l_C
\quad\textrm{or equivalently}\quad
[x,\tl{p}=mc\,\eta]=\, i\,\hbar.
\ee

\medskip

Before moving on understanding the physics of this prescribed non-commutative space, let us point out the possibility of a non-isotropic choice of basis of the coordinate operators. For example, there is an equivalent of the $\kappa$-Minkowski basis of DSR. Let us choose a particular direction, say $x_1$. Then we can define a new basis:
\be
\left|
\begin{array}{ccl}
\tl{x}_1&=& x_1, \\
\tl{x}_2&=&x_2+l_CL_{12}, \\
\tl{x}_3&=&x_3+l_CL_{13}.
\end{array}
\right.
\ee
These coordinate satisfy the interesting algebra:
\be
[\tl{x}_1,\tl{x}_a]=\,il_C\,\tl{x}_a,\qquad
[\tl{x}_2,\tl{x}_3]=0.
\ee
Although these commutators are indeed invariant under 3d rotations, these coordinates are not strictly speaking isotropic, since the three space coordinates do not transform the same way under rotations. Nevertheless, it could be considered as coordinates relative to an observer moving in the $x_1$ direction. Indeed, he would be unable to measure precisely the $x_1$ coordinate of systems (since he can not know his own position for sure), but then the two other coordinates, $\tl{x}_2$ and $\tl{x}_3$, would commute. However such coordinates are not directly useful in the framework of Special Relativity, so that we we will not focus on them. Nevertheless, we will shortly discuss them in the section \Ref{RNCG}.

\section{Physics of Special relativity from the non-commutative point of view}

In this section we address the physical interpretation of the non-commutative structure we introduced. Interestingly we face exactly the same type of ambiguities that arise in DSR. We have however here 99 years of  experiments and understanding of Special Relativity which allow us to give a clear interpretation and resolution of these ambiguities.

We start by  defining the deformed speed composition, which can be seen arising from a co-product structure.
This deformed product appears to be both non-commutative and non-associative. These features also found in DSR are sources of many debates in that context. Nevertheless, we are able to understand it in the present framework of Special Relativity as linked to a well-known physical phenomenon, the {\it Thomas precession}. The physical meaning of the choice of the speed coordinates is analyzed. This also a source of big debates in DSR, and we show how physics here allows to settle down this mathematical issue. We continue by defining the new notion of energy and momentum. Indeed as the speed space is deformed, those objects take a new shape. We define then the new co-product on  the galilean momenta describing the scattering of particles/systems. This co-product is defined accordingly to the principle of relativity, such that the conservation laws of energy and momenta are respected in all reference frames.

\subsection{Speed composition}

We have described the geometrical structure of the speed space. We will now derive the resulting new law of composition of speeds and check that there is indeed a maximum allowed speed. Let us insist that the speed composition is essentially different than the scattering of particles i.e addition of momenta.

We reconstruct the speed addition from the group structure we used. Indeed defining the speed space as the coset $\SO(3,1)/\SO(3)$, we have identified speeds with the boost sector. In order to keep calculations as simple as possible, we choose the spinorial representation of the Lorentz group. 4-vectors are represented as $2\times 2$ Hermitian matrices:
\be
(\alpha_0, \v{\alpha})\quad\longrightarrow\quad
\mat{cc}{\alpha_0+\alpha_3 & \alpha_1+i\alpha_2 \\ \alpha_1-i\alpha_2 & \alpha_0-\alpha_3}.
\ee
Defining the Pauli matrices $\v{J}$ as
$$
J_1=\left(\begin{array}{cc} &1 \\1& \end{array}\right),
\; J_2=\left(\begin{array}{cc} &-i \\i & \end{array}\right),\;
J_3=\left(\begin{array}{cc} 1& \\&-1 \end{array}\right),
$$
Lorentz group elements are the $U=\exp(\v{u}\cdot\v{J})$ for arbitrary complex vectors $\v{u}$. More precisely, boosts are parametrized as:
$$
g= e^{\frac{\eta}{2} \overrightarrow{b}\cdot\overrightarrow{J}}= \cosh\frac{\eta}{2} Id+ \sinh\frac{\eta}{2}
\overrightarrow{b}\cdot \overrightarrow{J},
$$
where $\eta\in\R$ is the boost parameter (or hyperbolic angle) and $\vec{b}$ is a normalized 3-vector indicating the boost direction. The group elements act on $2\times 2$ matrices by conjugation. Therefore, defining the origin of the hyperboloid as the 4-vector $V_0= (1,0,0,0)$ or equivalently as the $Id$ matrix, boosts allow to translate it to any point on the hyperboloid:
\be
V(\eta)\equiv\, g V_0 g^\dagger \,=\,(\cosh \eta, \sinh\eta \overrightarrow{b}).
\ee
We choose to proceed in the Snyder basis using the standard relativistic speed coordinates, but we bear in mind that the construction can be performed in any coordinate system. We shall come back to this issue in the next subsection. In the following, the speed  is therefore given by $\v{v}= c\, \v{\alpha}/\alpha_0 = c\, \tanh \eta \overrightarrow{b}$. Considering two speeds $\v{v_1},\v{v_2}$ and the corresponding boosts $g_1,g_2$, the composed speed $\v{v_1}\oplus \v{v_2}$ is defined as corresponding to the group element $g=g_1g_2$. This composition represents a change of reference frame: we are looking at a system moving with the speed $\v{v_2}$ in a reference frame moving at a speed $\v{v_1}$ in our reference frame. Mathematically, it corresponds to translating the point on the hyperboloid corresponding to the system 2 by the boost $g_1$. After some lengthy calculations that can be found in the appendix, we get the general formula for the composition of speeds:
\begin{equation}\label{speedcompr}
\overrightarrow{v_1}\oplus \overrightarrow{v_2}= \frac{1}{1+\frac{\overrightarrow{v_1}.\overrightarrow{v_2}}{c^2}}
\left(\overrightarrow{v_1}+ \frac{1}{\gamma_1}\overrightarrow{v_2}
+\frac{1}{c^2}\frac{\gamma_1}{1+\gamma_1}(\overrightarrow{v_1} .\overrightarrow{v_2})\overrightarrow{v_1}\right).
\end{equation}
When  $\overrightarrow{v_1}$ and $\overrightarrow{v_2}$ are collinear, it obviously reduces to the well-known formula for speed composition in Special Relativity:
\begin{equation}
\overrightarrow{v}=\frac{\overrightarrow{v_1}+\overrightarrow{v_2}}{1+
\frac{1}{c^2}\overrightarrow{v_1}.\overrightarrow{v_2}}.
\end{equation}
It is easy to check that the maximal speed is indeed $c$. Interestingly, when the speeds, $\overrightarrow{v_1}$ and
$\overrightarrow{v_2}$ are not collinear, the composition is non-commutative and moreover non-associative. This can be easily seen  at the level of the group element products. First, as matrices, two boost $g_1$ and $g_2$ won't commute (generically). Even more important, we are working on the coset $\SO(3,1)/\SO(3)$ which is a homogeneous space but not a Lie group. Indeed the group product of the two boosts is not a boost but a generic Lorentz transformation. Using the Cartan decomposition of a Lorentz transformation, we can write $g_1g_2=h_{12}g_{12}$, where $h_{12}$ is a (space) rotation and $g_{12}$ a pure boost. The composed speed $\v{v_1}\oplus\v{v_2}$ is extracted from the boost $g_{12}$ but the rotation $h_{12}$ does have a mathematical and physical impact. As shown in the appendices, it is expressed as
$$
h_{12}=e^{i\f{\theta}{2}\v{r}.\v{J}}=\cos\f{\theta}{2}+i\sin\f{\theta}{2}\v{r}.\v{J}
\quad\textrm{with}\quad
\left(\tan\f{\theta}{2}\right) \v{r} =\frac{1}{1+
c^{-2}\overrightarrow{v_1}.\overrightarrow{v_2}}(\overrightarrow{v_1}\wedge\overrightarrow{v_2}),
$$
which shows that $h_{21}=h_{12}^{-1}$. This rotation factor $h$ is indeed the (mathematical) reason for the non-associativity. Let us consider three boosts $g_i$. Doing $(g_1g_2)=h_{12}g_{12}$ and then $g_{12}g_3=h_{123}g_{123}$, or $(g_2g_3)=h_{23}g_{23}$ and then $g_1g_{23}=h'_{123}g'_{123}$ will generically give us  two different boosts $g_{123}$ and $g'_{123}$ so that  associativity will be violated. More explicitly\footnote{It is important to keep in mind that if $h$ is a rotation and $g$ a pure boost, then $hgh^{-1}$ is still a pure boost.}:
$$
(g_1g_2)g_3=g_1(g_2g_3)
\,\Rightarrow\,
h_{12}g_{12}g_3=h_{23}(h_{23}^{-1}g_1h_{23})g_{23}
\,\Rightarrow\,
h_{12}h_{123}g_{123}=h_{23}\tl{h}_{123}\tl{g}_{123}.
$$
This way, we see that although $(\v{v_1}\oplus\v{v_2})\oplus\v{v_3}$ (given by $g_{123}$) is different from $\v{v_1}\oplus(\v{v_2}\oplus\v{v_3})$ (given by $g'_{123}$), a modified associativity law still holds:
$$
(\v{v_1}\oplus\v{v_2})\oplus\v{v_3}=(h_{23}\vartriangleright\v{v_1})\oplus(\v{v_2}\oplus\v{v_3}).
$$
The non-commutativity and non-associativity have an interesting physical manifestation as the so-called {\it Thomas precession}\cite{thomas}, so that we actually name the rotation $h$ the Thomas precession (associated to the change of reference frame). A simple application of the phenomenon deals with a particle moving around a circular path. The particle undergoes continuous acceleration, but at each instant it is at rest with respect to the momentarily co-moving inertial frame. If we consider the "parallel transport" of a vector around the continuous cycle of momentary inertial rest frames of the particle, we find that the vector does not remain fixed. Instead, it "precesses" as we follow it around the cycle. This relativistic precession (which has no counter-part in non-relativistic physics) actually has observable consequences in the behavior of atomic particles. This precession is to be thought of as a holonomy created because of the relativistic effect.

\medskip

In the galilean context, the speed space was a (abelian) group, when going to the relativistic case, i.e. implementing the maximal speed, one seems to completely lose this structure since the product now is non-associative. In fact, it is simply that we are working on a group coset $G/H$.  The resulting abstract structure can be called {\it  gyro-group}  or {\it gyro-vector space} (as in \cite{ungar}), or quasi-group (as in for example \cite{3cocycle}). We have modified notions of commutativity ("gyro-commutativity") and associativity ("gyro-associativity"):
\begin{itemize}
\item{\it Gyrocommutativity} \begin{equation} \label{gyrocom}
u. v= Ad(h_{uv})(v. u) \end{equation}
\item{\it Gyroassociativity} \begin{equation}\label{gyroasso}\begin{array}{c}
u.(v. w)= (u. v).Ad(h_{uv}) w \\
(u. v). w= u. (v . Ad(h_{vu})w )
 \end{array}\end{equation}
\end{itemize}
where $u,v,w$ are group elements in the coset $G/H$, the $h$'s element of the subgroup $H$ and $Ad$ the adjoint action of $H$ on the elements of the coset. It is very intriguing to understand  what is the dual algebraic structure which
describes this gyrostructure. One might think of a quantum group structure with a quasi-triangular structure given by a $R$ matrix, but it seems more complicated. We shall come back on this non-commutative structure in a section below.


\medskip

Let us then describe the dual structure to the change of reference frames. Let us note the space of deformed speeds $\vv$. The speed composition du change of frame is given by the map ($r$ standing for relativistic):
\begin{equation}\label{ff}
\begin{array}{rccl} \ff:&\vv\otimes \vv &\rightarrow &\vv \\
&(\overrightarrow{w},\overrightarrow{v})&\rightarrow& \overrightarrow{w}\oplus_r\overrightarrow{v},
\end{array}\end{equation}
Dually we can consider  two frames, and the associated state  $\ket{1,2}$ which represents the frame 1 associated to
the observer seen in the frame 2. With respect to the speed operator, we therefore have a co-product that can be
determined such that
$$
\cop_{\ff}\overrightarrow{V}\ket{1,2}= (\overrightarrow{v_1}\oplus \overrightarrow{v_2})\ket{1,2}.
$$
In the galilean case, we just have $\overrightarrow{v_1}\oplus_{g} \overrightarrow{v_2} = \overrightarrow{v_1}+
\overrightarrow{v_2}$, so that $\cop_{\ff}\overrightarrow{V}=\overrightarrow{V}\otimes 1+ 1 \otimes\overrightarrow{V}$.
In the Special relativity case, the co-product is more involved as $\overrightarrow{v_1}\oplus_{r}
\overrightarrow{v_2}$ follows from the law (\ref{speedcompr}).

Let us insist on the central point that this coproduct must not be confused with the coproduct describing the
addition of momenta when thinking of scattering of particles: they are {\it essentially} different. Note that this non-triviality of the co-product $\cop_{\ff}$ is analogous to the non-triviality of the  coproduct $\cop$  that arises in DSR. Indeed we are here deforming and curving the space of speed and therefore getting a non-trivial structure (sum rule) for them just as in DSR the space of momenta is deformed with a non-trivial structure. Going down to more details, one can actually state than in Special Relativity we deform/curve the momentum space on the mass-shell (at fixed rest mass $m$) while DSR goes further and deforms/curves the full momentum space off the mass-shell.

\medskip

Let us illustrate the coproduct construction the toy model given by the 2d case. Note that here the speeds are always collinear so that we don't face the non-associativity issue in this context. In terms of the boost parameter $\eta$ which is simply the hyperbolic angle,  the speed addition is very simply given by:
\begin{equation}
\eta_{tot}=\eta_1+\eta_2.
\end{equation}
The associated coproduct is then trivial
\begin{equation}
\Delta \eta = \eta\otimes 1  + 1\otimes \eta.
\end{equation}
Considering the Special Relativity coordinates, the relevant speed is $v=c\tanh \eta$.
The speed addition is now not trivial but can be derived. Indeed, starting from
\begin{equation}
v_1\oplus v_2 = c\tanh(\eta_1+\eta_2)= c\frac{\tanh\eta_1+\tanh\eta_2}{1-\tanh\eta_1\tanh\eta_2}= \frac{v_1+v_2}{1+
c^{-2}v_1v_2},
\end{equation}
one can construct the dual coproduct using the fact that $\frac{1}{1+X}\sim
1+X+X^2+X^3+\hdots$ and $\Delta X^n=(\Delta X)^n$, which combined lead to the formal equality $\cop(\frac{1}{1+X})=
\frac{1}{1\otimes1+\cop X}$:
\bes
\Delta_{\ff} v &=& c\, \left(\Delta_{\ff}  \tanh \eta\right) \nn\\
&=&  \frac{v\otimes 1 + 1\otimes v }{1 + c^{-2} v\otimes v}.
\ees
Once again, this coproduct for the momentum is not associated to the scattering of particles but to the behavior of the speed under change of reference frame.

\subsection{Which coordinate is physical?}
In the previous subsection we have chosen the Snyder basis and the standard the Special Relativity velocity. However nothing prevents us {\it a priori} to use another coordinate system and therefore another definition of the velocity. As a matter of fact, we could have for example defined the speed as $\v{v}\equiv\tanh\frac{\eta}{2}\v{b}$ (let's recall that $\v{b}$ is a normalized 3-vector). If we had done so, the speed composition would have been constructed the same way and derived from the group multiplication, but we would have gotten a different final result. Indeed one can see that in this particular case the result is
\begin{equation}
\overrightarrow{v}= \frac{\left(
2-\frac{1}{\gamma_{v_2}^2}+\frac{2}{c^2}\overrightarrow{v_1}.\overrightarrow{v_2}\right)\overrightarrow{v_1}+
\frac{1}{\gamma_{v_1}}\overrightarrow{v_2}}{1+ \frac{2}{c^2}\overrightarrow{v_1}.\overrightarrow{v_2} +
\frac{1}{c^4}v_1^2v_2^2}.
\end{equation}
One could also have chosen $|\v{v}|\equiv\sinh\eta$, and we shall see alter that this choice actually corresponds
to the standard relativistic speed ($\gamma \v{v}$).  In fact one can generalize the construction for any function $f$ defining the speed in terms of the rapidity $cf(\eta)=|\overrightarrow{v}|$. $f$ encodes the map between the Galilean space of velocities $\R^3$ and the curved space of velocities defined by the hyperboloid. Such a map can be understood as a map between (boost) Lie algebra elements parametrized by $\v{v_g}=\eta \v{b}$ and (boost) Lie group elements (such a map is not unique since it depends on the coordinate system chosen on the Lie group manifold). Making the common sense assumption that physics is isotropic, so that the deformation map should only be a function of the norm of the speed and not of its direction, the map can be defined as $\v{v}=c\varphi(\eta \v{b})=cf(\eta)\v{b}$.
We only require that $f:\R_+\arr\R_+$ be a continuous increasing function, onto $\R_+$ or a compact interval $[0,K]$.
For an arbitrary function $f$, one can derive the general formula:
\begin{equation}\label{speedcomp}
\overrightarrow{v}= \overrightarrow{v_1}\oplus\overrightarrow{v_2}= \frac{f(\eta_v)}{\sinh
\eta_v}(C_{v_1}\overrightarrow{v_1}+C_{v_2}\overrightarrow{v_2}),
\end{equation}
with
\begin{equation}\label{cvi}
\begin{array}{rcl}
C_{v_1}&=& \frac{\sinh\eta_{1}}{f(\eta_{1})}\cosh\eta_{2} +\left( \frac{\sinh\eta_{1}}{f(\eta_1)}\right)^2\frac{\sinh\eta_{2}}{f(\eta_2)}\frac{1}{1+\cosh\eta_{1} }\frac{\overrightarrow{v_1}.\overrightarrow{v_2}}{c^2}\\
C_{v_2}&=& \frac{\sinh\eta_{2}}{f(\eta_{2})}
\end{array}
\end{equation}
We have also the quantity (which we shall see is the new definition of energy)
\begin{equation}\label{energycomp}
\cosh \eta_{v}= \cosh \eta_{1}\cosh \eta_{2}+ \frac{\sinh \eta_{1}}{f(\eta_{1})}\frac{\sinh
\eta_{2}}{f(\eta_{2})}\frac{\overrightarrow{v_1}.\overrightarrow{v_2}}{c^2}.
\end{equation}

\medskip

As we see, there is an ambiguity in what is to be called the velocity. The only way to overcome this issue is physical experiments. Effectively measuring and experimentally testing the composition of speeds (and it is enough to study the composition of collinear velocities) will specify which deformation function $f$ needs to be chosen. Let us point out that the original Galilean case corresponds to the trivial choice $f=Id, v=\eta$. Choosing different $f$'s leads to a whole family of mathematically consistent relativistic theories with bounded speeds.

Another experiment to extract the right function is the measure of length contraction. Physics on the hyperboloid tells us that the length will be contracted by a factor $\gamma\equiv\cosh\eta$ for an observer with velocity $v/c=f(\eta)$. Measure of the contraction factor $\gamma$ in terms of the velocity will then specify us which is the function realized in Nature.
As an example, if the speed was given by $v= c\tanh\frac{\eta}{2}$, then the contraction factor $\gamma = \cosh\eta$ as a function of $v$ would be:
\begin{equation}
\gamma= \frac{1+\frac{1}{c^2}v^2}{1-\frac{1}{c^2}v^2}
= 1 + 2\f{v^2}{c^2}+2\f{v^4}{c^4}+\hdots
\end{equation}
Whereas we know (through experiments) that we should get:
\begin{equation}
\gamma= \frac{1}{\sqrt{1-\frac{1}{c^2}v^2}}= 1+ \frac{1}{2c^2}v^2-\frac{3}{8c^4}v^4+\hdots
\end{equation}
which leads to the usual relativistic definition $v= c\tanh\eta$.

To summarize, it is understood that the "true physics" is described by a single system of coordinates on the hyperboloid leading to a unique choice of deformation: although there is a whole class of equally consistent deformed speed additions, physics should select a unique deformation. The way to find the right one is to  use experiments to calibrate to the right deformation. Once this is done, we can proceed to actually do {\it predictions}. Once the theory calibrated, agreement with further experiments will really test the theory and overall consistence between the different experiments will improve the precision of the calibration.

This single system of coordinates define the physical meaning of the mathematical entities or equivalently defines what are the measured physical quantities like what we call the energy or spatial coordinates. Of course, we can always use arbitrary systems of coordinates to describe the physics and phenomena, as long as we remember the definition of the actual physical quantities. In other words, we can still do passive coordinate change without affecting the underlying physics.

An immediate prediction that one can do here is the Thomas precession, and its precise value depends on the actual definition of the velocity. It is interesting to note that this experiment is precisely showing the non-associativity of the speed composition which is something somehow unexpected, as associativity is usually taken as granted.

To introduce more complex experiments, we need further tools and define the other physically relevant quantities such as energy and momentum, which we will now discuss.

\subsection{New definition of energy and momentum}

As we have deformed the speed addition, it is expected that the usual notions of momenta and energy also be modified.
The Galilean energy and momentum are easily expressed in term so the velocity and the mass (of the system/particle):
\be
e\equiv\demi mv^2=\demi \frac{p^2}{m}, \qquad \overrightarrow{p}\equiv m\overrightarrow{v}.
\ee
Momenta generates translation in space, while the energy, considered as the Hamiltonian, generates the motion of the particle/system. In the Galilean context, we indeed have:
$$
\{p_i,x_j\}_{(g)}=\,\delta_{ij},
\qquad
\{e,x_i\}_{(g)}=\,\f{p_i}{m}=v_i.
$$
Now, in the relativistic framework, assuming the modified Poisson bracket structure \Ref{poisson}, we have:
\be
\{p_i,x_j\}=\,\delta_{ij}-\f{v_iv_j}{c^2}=\,\delta_{ij}-\f{p_ip_j}{m^2c^2},
\qquad
\{e,x_i\}=\,\f{p_i}{m}\left(1-\f{|v|^2}{c^2}\right)=\,v_i\left(1-\f{v^2}{c^2}\right).
\ee
An important point is that these relations are not linear anymore, so that it doesn't provide us with a Lie algebra structure. If we want to identify the Lie algebra generating the translations and motion of the system, we need to go back to the $\alpha_\mu$ variables. Indeed, we recall the definition of our framework:
$$
[x_i,\alpha_0]=\,i\f{\hbar}{mc}\alpha_i,
\qquad
[x_i,\alpha_j]=\,i\f{\hbar}{mc}\alpha_0\delta_{ij}.
$$
These relations are linear and define a Lie algebra structure. It is therefore natural to define the new notion of a relativistic energy-momentum through the $\alpha_\mu$ by giving them the right units:
\be
E=\,\alpha_0\,mc^2=\,\gamma mc^2,
\qquad
P_i=\,\alpha_i\,mc=\,\gamma p_i,
\ee
where we are using the usual relativistic factor $\gamma\equiv \alpha_0$.
These new energy and momenta define the following algebra:
\be
[x_i,E]=\,i\hbar\f{P_i}{m},
\qquad
[x_i,P_j]=\,i\hbar\f{E}{mc^2}\delta_{ij},
\label{EP}
\ee
where we notice that we have $\{x_i,P_j\}=\delta_{ij}$ only up to an energy-dependent factor $E/mc^2=\gamma$. This factor is equal to 1 when and only when the velocity $v$ is equal to 0, then increases and diverges to $\infty$ when the velocity reaches the maximal speed, or speed of light, $c$.

The reason why linearity is important is when considering composite systems. Indeed for {\it free systems}, energy and momentum are assumed to be {\it extensive quantities}. This generically requires linearity  of the bracket between position coordinates and energy-momentum. Indeed, using the relativistic energy-momenta $E^{(i)},\v{P}^{(i)}$, then the total energy-momentum of the composite system is naturally simply:
$$
E^{(tot)}=E^{(1)}+E^{(2)}, \qquad
\v{P}^{(tot)}=\v{P}^{(1)}+\v{P}^{(2)}.
$$
To check this, we consider the coarse grained position operators which are simply the center of frame coordinates:
$$
X\equiv \f{m_1x^{(1)}+m_2x^{(2)}}{M},
$$
where the coarse grained mass is naturally defined as $M\equiv m_1+m_2$.
Then it is easy to verify that the following commutators hold:
\be
[X_i,E^{(tot)}]=\,i\hbar\f{P_i^{(tot)}}{M},
\qquad
[X_i,P_j^{(tot)}]=\,i\f{\hbar}{c^2}\f{E^{(tot)}}{M}\delta_{ij},
\ee
and we indeed recover the same algebra structure \Ref{EP}.
At this point we must point out that there is no constraints on the definition of the coarse grained mass. However, we must require that the total energy-momentum be on the $M$ mass-shell hyperboloid. This lead us to a modified
total mass:
\be
M^2\equiv E^2-\v{P}^2= m_1^2+m^2_2+2m_1m_2\gamma_1\gamma_2\left(1-\f{\v{v_1}.\v{v_2}}{c^2}\right).
\ee
It is straightforward to check that $\gamma_1\gamma_2(1-\v{v_1}.\v{v_2}/c^2)\ge1$, so that the Galilean total mass $m_1+m_2$ is not the coarse-grained mass anymore but simply a lower bound for it.

\medskip

Now that we have seen that the relativistic energy-momentum is simply additive for composite systems as expected in a kinematical theory for free systems, we can look at the situation from the point of view of the Galilean energy-momentum and interpret the velocity bound as inducing (or eventually induced by) an interaction potential between the two systems considered in the Galilean context. Indeed, the composite system is relativistically described by the total mass $M$ and the velocity  of the center of frame given by
$$
\v{v}=\,c^2\f{\v{P}^{(tot)}}{E^{(tot)}}=\, \f{\gamma_1m_1\v{v_1}+\gamma_2m_2\v{v_2}}{\gamma_1m_1+\gamma_2m_2},
$$
so that the composite Galilean energy is:
\be
e^{(tot)}=\,\f{1}{2}M\v{v}^2
=\f{1}{2}\,
\f{|\gamma_1m_1\v{v_1}+\gamma_2m_2\v{v_2}|^2}{(\gamma_1m_1+\gamma_2m_2)^2}
\sqrt{m_1^2+m^2_2+2m_1m_2\gamma_1\gamma_2\left(1-\f{\v{v_1}.\v{v_2}}{c^2}\right)}\, .
\ee
It is rather obvious that $\Delta e\equiv e^{tot}-(e^{1}+e^{2})$ does not vanish, and we can interpret this difference as an interaction potential $V\equiv \Delta e$ between the two systems depending on their velocities. This potential would forbid the velocities to exceed the bound $c$.

\medskip

The most important point about energy-momentum is its conservation. We will see below that this provide us with another argument why we need new relativistic energy-momenta instead of the old Galilean notions. The key point is actually the {\it relativity principle}: If the energy-momentum is conserved (during a process/scattering) in one reference frame, it should be seen as conserved in {\it any} reference frame. More precisely, considering the scattering of two particles/systems of given (rest) masses and of speeds $\v{v_1},\v{v_2}$, the relativity principle can be translated to the equivalence relation:
\be
E_{v_1}+E_{v_2} \textrm{ and } P_{v_1}+P_{v_2} \quad \textrm{conserved}\quad
\Leftrightarrow \quad
\forall\, \v{v},\quad
E_{v\oplus v_1}+E_{v\oplus v_2} \textrm{ and } P_{v\oplus v_1}+P_{v\oplus v_2} \quad \textrm{conserved}.
\ee
We will see that the Galilean energy-momentum do not satisfy this criteria anymore when considering the relativistic composition of speeds, but that the newly defined relativistic quantities will do so, so that they will become physically meaningful.

\subsection{Scattering, Conservation law and Principle of Relativity}

In this section we address the issue of scattering (i.e. coproduct) and  conservations laws (compatibility/consistency between the different coproducts attached to different reference frames). Generally, considering a scattering process between two systems 1 and 2, the total energy-momentum of the composite system $1+2$ is conserved. In the Galilean framework, the addition of energy-momentum is trivial:
$$
\v{p}^{tot}=\v{p}^1+\v{p}^2, \qquad
e^{tot}=e^1+e^2,
$$
which can be expressed in terms of the dual coproduct structure:
$$
\cop \overrightarrow{p}= \overrightarrow{p}\otimes 1 +1 \otimes \overrightarrow{p},\qquad
\cop e= e\otimes 1 +1 \otimes e.
$$
As seen above, consistency with the velocity bound requires to change to a new notion of relativistic energy-momentum:
\be
E\equiv\alpha_0\,mc^2=\,mc^2\cosh\eta=\,\gamma mc^2,
\ee
\be
\v{P}=\alpha_i\, mc=\,mc\sinh\eta \,\v{b}=\,\gamma mc\tanh\eta\,\v{b}=\,mc\f{\sinh\eta}{f(\eta)}\,\v{v},
\ee
where the speed is defined through the deformation function $f$ as $\v{v}=f(\eta)\v{b}$. An important point is that our definition of energy-momentum is actually independent of the deformation function $f$. Now, in the new framework with bounded speed, it is the new notion of energy-momentum which has a trivial coproduct:
$$
E^{tot}=E^{1}+E^{2}, \qquad
\v{P}^{tot}=\v{P}^{1}+\v{P}^{2}.
$$
An useful exercise to realize how much Special Relativity is not natural from the Galilean point of view is to compute the coproduct on the Galilean energy-momentum in the new deformed framework. Choosing the usual $f=\tanh$, we have:
\be
\label{addrelmomenta}
\v{P}=m\gamma\v{v}=\v{P_{1}}+\v{P_{2}}=m_1\gamma_1\overrightarrow{v_1}+m_2\gamma_2\overrightarrow{v_2},
\ee
\be
\label{addrelenergy}
E =\gamma m= \gamma_1 m_1+ \gamma_2 m_2.
\ee
For this, we can deduce the values of the speed $\v{v}$ and the mass $m$:
$$
\v{v}=\f{\v{P}}{E}=
\f{m_1\gamma_1\v{v_1}+m_2\gamma_2\v{v_2}}{\gamma_1 m_1+ \gamma_2 m_2},
$$
$$
m=\f{\gamma_1 m_1+ \gamma_2 m_2}{\gamma},
$$
where $\gamma$ is simply defined as $\gamma^{-2}=1-v^2/c^2$. Thus we get:
\be
\v{p}=\f{1}{\gamma}\overrightarrow{P}
=\frac{1}{\gamma} (\gamma_1\overrightarrow{p_{1}}+\gamma_2\overrightarrow{p_{2}}),
\ee
\be
e= \f{1}{2\gamma}\f{|\v{P}|^2}{E}
=\f{1}{2\gamma}\f{|\gamma_1\v{p_1}+\gamma_2\v{p_2}|^2}{\gamma_1 m_1+ \gamma_2 m_2}.
\ee
Substituting in these expressions $m_i=|\v{p_i}|^2/e_i$ and $\gamma_i^{-1}=\sqrt{1-4e_i^2/p_i^2c^2}$ everywhere needed, we get the expression of $e_{tot}$ and $\v{p}_{tot}$ in terms of $e_1,e_2,\v{p_1},\v{p_2}$. This formula is much more complicated than in the Galilean theory. And moreover this coproduct now mixes the energy $e$ and the momentum $\v{p}$ so that we need to work anyway with the full Galilean energy-momentum quadri-vector.
Explicitly deriving the full exact coproduct is pretty cumbersome, and fortunately not really needed here.

\medskip

Let us point out the new energy-momentum naturally satisfy a new dispersion relation different from the initial Galilean one:
\be
p^2=2me
\quad\longrightarrow\quad
E^2=P^2c^2+m^2c^4.
\ee
Let us recall how to go from the relativistic relation to the Galilean one. Indeed we have
$P^2c^2=(E-mc^2)(E+mc^2)$. Noticing that we always have $E\ge mc^2$, we can define $\tl{E}=E-mc^2$. And when the renormalized energy $\tl{E}$ is close to 0 i.e $E\approx mc^2$, we can approximate the dispersion relation by $P^2c^2=2\tl{E}mc^2$.

\medskip

Let us now explain why the addition of relativistic speed is the trivial one. One of the basic postulates of physics is the principle of relativity stating that different observers should still experiment the same laws of physics. More precisely, we require to have the same laws of conservation in any reference frames. More technically this means that we want a compatibility relation between the coproduct describing the scattering process (i.e going from two systems to one coarse-grained composite system) and the coproduct describing the change of (reference) frame.
Noting $\oplus_w$ the change of reference frame with speed $\v{w}$, we need to check that for all $\v{w}$ we indeed have $(\oplus_w\otimes\oplus_w)\circ\cop=\cop\circ\oplus_w$.
More explicitly, we consider two systems $m_1,\v{v_1}$ and $m_2,\v{v_2}$, and another reference frame defined by its (arbitrary) speed $\v{w}$ with respect to the initial frame. We require that the conservation of the total energy momentum of the composite system $1+2$ in the initial frame be exactly equivalent to its conservation in the moving frame: we impose that the total energy-momentum of
$(m_1,\v{v_1})+(m_2,\v{v_2})$ is conserved if and only if the total energy-momentum of
$(m_1,\v{w}\oplus\v{v_1})+(m_2,\v{w}\oplus\v{v_2})$ is conserved.

It is interesting to note that the order in which one composes the speeds is very important here. Indeed
the composition of velocities is not abelian anymore in general. We want to consider the systems, 1 and 2, and translate them in the momentum space, and not take a single system, defined by the speed $w$, and translate it in two arbitrary directions 1 and 2. This means we are truly considering the addition $\v{w}\oplus\v{v_i}$, and not
$(\overrightarrow{v_i}\oplus \overrightarrow{w})$. In fact the latter choice gives completely unrealistic results with respect to the energy-momentum conservation law\footnote{It tells us that the rest mass of the two particles should be conserved during any process, so that this restricts too much the possible interactions consistent with the kinematical framework. }.
Finally, the trivial coproduct on the modified energy-momenta is a solution of a such compatibility relation, as we are checking below. A priori it should be the only one but we haven't proven this.

\medskip

First let's have a look at the energy,
\begin{equation}
E\equiv\gamma mc^2 =E_1+E_2=\gamma_1 m_1c^2+ \gamma_2 m_2c^2
\quad\longrightarrow\quad
E'\equiv \gamma_1' m_1c^2+ \gamma_2' m_2c^2,
\end{equation}
with $\gamma_i=\cosh\eta_{v_i}$ and $\gamma'_i = \cosh\eta_{v_i+w}$. Computing $\gamma'_i$ from equation (\ref{energycomp}), we get the formula:
\bes
E'&=& (m_1c^2\cosh \eta_{1}+ m_2c^2\cosh \eta_{2} )\cosh \eta_{w}+ \left(m_1\frac{\sinh
\eta_{1}}{f(\eta_{1})}\overrightarrow{v_1}+ m_2 \f{\sinh \eta_{2}}{f(\eta_{2})} \overrightarrow{v_2}\right).
\f{\sinh \eta_{w}}{f(\eta_{w})} \overrightarrow{w}\nn\\
&=& (E_1+E_2)\cosh \eta_{w} + \f{\sinh \eta_{w}}{f(\eta_{w})}\v{w}.\left(\v{P_1}+\v{P_2}\right) \nn\\
&=& E\,\cosh \eta_{w} + \f{\sinh \eta_{w}}{f(\eta_{w})}\v{w}.\v{P}
\label{E'}
\ees
In the same way, the total relativistic momentum reads:
\begin{equation}
\overrightarrow{P}=\overrightarrow{P_{1}}+\overrightarrow{P_{2}}
\quad\longrightarrow\quad
\overrightarrow{P}'=\overrightarrow{P_{1}}'+\overrightarrow{P_{2}}'.
\end{equation}
Once again we apply the rule of speed composition, as given in \Ref{speedcomp}. Now defining $\v{w}=cf(\eta_w)\v{b}$, a trick to compute the change in momenta due to the change of reference frame is to remember that $\v{P_i}=mc\sinh\eta_i\,\v{b_i}$, so that a fast way to get the momentum in the moving frame is to apply the speed composition \Ref{speedcomp} with the new deformation function $\tl{f}\equiv\sinh$. This way, we get:
\be
\f{\v{P_i}'}{m_ic}=
\left(\cosh\eta_1+\f{1}{1+\cosh\eta_w}\sinh\eta_1\v{b_1}.\sinh\eta_w\v{b}\right)\,\sinh\eta_w\v{b}
+\sinh\eta_1\,\v{b_1}.
\ee
We can then easily deduce that:
\be
\label{P'}
\v{P}'=\v{P}+\f{\sinh\eta_w}{f(\eta_w)}\v{w}\,
\left(\f{E}{c}+\f{\sinh\eta_w}{f(\eta_w)(1+\cosh\eta_w)}\v{w}.\v{P}\right).
\ee

\medskip

Having fixed the change of frame $\v{w}$, the equations \Ref{E'} and \Ref{P'}, giving the energy-momentum in the moving frame in terms of the original energy-momentum, show that the conservation of the modified energy-momentum $(E,\v{P})$ in one frame is equivalent to its conservation in any reference frame (moving at a constant speed with respect to the initial one). Let us insist on the fact that we need the full relativistic energy-momentum and we can not work only with the energy or only with the momentum. At the end of the day, considering these new notions of energy and momentum, the deformed Galilean theory still respects the principle of relativity. Moreover this result hold whatever the chosen deformation $f$, so that the principle of relativity does not select a particular deformation, and all these theories are both mathematically consistent and physically realistic. Only experiments allows us to select the specific deformation function $f(\eta)=\tanh\eta$. In that particular case, let us re-write the transformation of the energy-momentum under change of frame:
\bes
E'&=&\gamma_w(E+ \overrightarrow{P}.\overrightarrow{w})\nn\\
\overrightarrow{P}'&=& \overrightarrow{P}+ \v{w}\left(\gamma_w\f{E}{c}+ \frac{\gamma_w^2}{1+\gamma_w}\overrightarrow{w}.\overrightarrow{P}\right)
\ees
Let us underline that writing the relativistic laws of conservation of energy-momentum in terms of the initial Galilean notions $(e,\v{p})$ would take a very ugly aspect.

\medskip

To conclude, in this section we have shown that there are many possible deformations of the Galilean framework in order to accommodate a maximal speed. Then although all these deformed theories are mathematically consistent and can be related to each other simply through a change of coordinate systems on the hyperboloid, physics is really described by a unique deformation, which has to be calibrated through experiments.  The important thing in the end is that deforming the speed space and the Galilean symmetry forces to define new notions of energy and momenta in order to respect the principle of relativity.

A last point we would to insist upon is that the deformation comes together with a non-associativity, which might be unexpected but has a true physical meaning. This is linked with the fact that the non-commutative structure is somehow observer dependent, therefore in some sense "relative". We develop this point in the following section.

\section{Quantum Groups and the Notion of a Relative Non-Commutative Geometry}
\label{RNCG}

\subsection{Between Quantum Groups and Quasi-Groups}

The non-commutative and non-associative structure of Special Relativity comes from using a  curved space the
hyperboloid $\SO(3,1)/\SO(3)$ as configuration space. More precisely, the non-commutativity comes from using a
non-abelian group $\SO(3,1)$, and the non-associativity comes from using a quotient space which is a homogeneous space
but not a group manifold. Nevertheless, as we have seen, we have an exact control on both the non-commutativity and the
non-associativity, which is quantified through the Thomas precession. This has lead to the deformed notions of
gyro-commutativity and gyro-associativity. Then a natural question is whether such gyro-groups can be considered as
quantum groups. On the other hand, gyro-groups are understood to be {\it transversal} loops which are a special case of
{\it loops} which are themselves a particular case of {\it quasi-groups} (for a short review of these notions, see
\cite{quasigroups}).

The simplest case of gyro-groups is a quotient $G/H$ where elements of $H$ are just phases.  Then the group product
gives $g_1g_2=\exp(i\theta_{12})g_{12}$ while we define the product on the coset as $g_1.g_2=g_{12}$, and one can check
that the new product is still associative. A straightforward example is provided by the Heisenberg-Weyl group. Indeed
because $[x,p]\propto Id$  commutes with $x$ and $p$, multiplying group elements of the type $\exp(\alpha x+\beta p)$
simply induces a phase. More generally, when $H$ is a subgroup included in the center of $G$, $G/H$ still has a group
structure and the product on the coset stays associative. This is similar to the first example of non-commutative
geometry, given by the Moyal deformation. In this case, one deforms the product of the algebra of functions on the
space by introducing a phase factor:
$$
x\star y=e^{i\theta (x,y)}y\star x.
$$
The phase $\theta (x,y)$ is generally a scalar function of $(x,y)$. And by construction,  the Moyal product $\star$ is
associative.

Quasi-groups can be more involved. Already in the case of phases, one can introduce  a non-associative extension of
$\U(1)$. Noting $g_1g_2$ the usual group product, we define very generally a new product
$g_1*g_2=\exp(i\alpha(g_1,g_2))g_1g_2$ without using any coset structure. Then it is straightforward to check that:
$$
g_1*(g_2*g_3)=e^{i\Delta(g_1,g_2,g_3)}\,(g_1*g_2)*g_3,
$$
where $\Delta(g_1,g_2,g_3)=\alpha(g_2,g_3)-\alpha(g_1g_2,g_3)+\alpha(g_1,g_2g_3)-\alpha(g_1,g_2)$ defines a 3-cocycle. This can be actually used to describe the physics of Dirac's monopole \cite{3cocycle}.

The next simplest cases of gyro-groups are already naturally non-associative.  As an example, consider the sphere
${\cal S}_2=\SU(2)/\U(1)$ with $\U(1)$ being the (abelian) (sub-)group of rotations along the $z$-axis. The coset is
parametrized as $g(n)=\exp(i\hat{n}.\hat{J})=\exp(i\theta(\sin\phi J_x-\cos\phi J_y)$ with
$n=(\sin\theta\cos\phi,\sin\theta\sin\phi,\cos\theta)\in{\cal S}_2$. $g(n)$ actually maps the point $n_0=(0,0,1)$ to
the point $n$. Then:
$$
g(n_1)g(n_2)=g(n_{12})e^{i{\cal A}(n_1,n_2)J_z},
$$
where ${\cal A}(n_1,n_2)$ is the area of the geodesic triangle on the sphere  with vertices $n_0,n_1,n_2$. Already
because $J_z$ doesn't commute with $J_x,J_y$, the product on the sphere $g(n_1).g(n_2)=g(n_{12})$ is not associative.

\medskip

The general case states that, noting the group product $g_1g_2=h_{12}g_{12}$ and  the coset product $g_1.g_2=g_{12}$,
the gyro-associativity reads:
$$
(g_1.g_2).g_3=(Ad(h_{23})\rhd g_1).(g_2.g_3).
$$
With the opposite convention $g_1g_2=g_{12}h_{12}$, we would get:
$$
g_1.(g_2.g_3)=(g_1.g_2).(Ad(h_{12})\rhd g_3).
$$
At this point, we would like to stress that the $h$ factor depends on the choice of section we use to describe the
coset $G/H$ i.e the map $G/H\arr G$, or equivalently the map $\lalg{g}\arr G/H$ used to parametrize the homogeneous
space. Indeed it seems that it is generically possible to absorb the factor $h$ in the definition of the section, such
that the product on the coset becomes associative! In our case where we study $\SO(3,1)/\SO(3)$, such a parametrization
is provided by the modified coordinate operators $\tl{x}_i$ introduced at the end of section \Ref{section1}. Because
the $\tl{x}_i$'s form a Lie (sub-)algebra, the product of group elements $\exp(iw_i\tl{x}_i)$ is stable and doesn't
need any rotational $h$ factor. This is possible because the $\tl{x}_i$ are not pure boost anymore but include a
rotational part: \bes\label{noniso}
\tl{x}_1&\equiv& x_1=l_C \,N_1, \nn\\
\tl{x}_2&\equiv& l_C \,(N_2+L_3), \nn\\
\tl{x}_3&\equiv& l_C \,(N_3-L_2).
\ees
Using the spinorial representation, we notice that:
$$
\tl{x}_2=\,2\,l_C\,\mat{cc}{0&1\\0&0}=\,2l_C\,J_+,\qquad
\tl{x}_3=\,2il_C\,J_+=i\tl{x}_3.
$$
Now we choose the following section for the coset:
\be
g\left(\v{w}\right)=g(w,\hat{w})=e^{i\f{1}{2l_C}w\tl{x}_1}e^{i\f{1}{2l_C}\hat{w}_a\tl{x}_a}.
\ee
Such group elements map (by the adjoint action) the origin $(1,0,0,0)$ to the points on the hyperboloid:
\bes
\alpha_0&=&\cosh w+\f{1}{2}(w_2^2+w_3^3)e^{-w} \nn\\
\alpha_1&=&-w_3 \nn\\
\alpha_2&=&-w_2 \nn\\
\alpha_3&=&-\sinh w+\f{1}{2}(w_2^2+w_3^3)e^{-w}.
\ees
Let us underline that the speed $\v{v}\equiv c\v{\alpha}/\alpha_0$ is not collinear to the vector $\v{w}=(w,\hat{w})$ and also that $w$ is not the boost rapidity $\eta$ so that the true (physical) speed $|\v{v}|$ is not given by $\tanh|\v{w}|$. Then using the identity:
$$
e^{iw\f{x_1}{l_C}}e^{iw_a\f{\tl{x}_a}{l_C}}e^{-iw\f{x_1}{l_C}}=e^{ie^{-w}w_a\f{\tl{x}_a}{l_C}},
$$
it is straightforward to check that: \be
g\left(\v{w}^1\right)g\left(\v{w}^2\right)=g\left(w^1+w^2,e^{\f{w^2}{2}}\hat{w}^1_a+\hat{w}^2_a\right). \ee This
derived law of composition of the $\v{w}$'s is much simpler than the laws of speed  composition described earlier, and
it is easy to check it is also associative. Nevertheless, they are equivalent as long as we express the velocity
$\v{v}$ in terms of the 4-vector $\alpha_\mu$. Therefore, such coordinates can be very useful to do calculations
although the physical meaning of the coordinate $\v{w}$ is not straightforward. Now however, having a associative
composition law contrasts with the earlier claim that non-associativity is physical and that the Thomas precession is
an experimental fact in the context of Special Relativity. This apparent paradox originates in the fact that the new
coordinates include a rotational part and do not have a straightforward physical meaning: the $\tl{x}$'s are not the
physical measured coordinates and the $\v{w}$'s are not the physical velocities. In order to go further, it should be
interesting to analyze/describe the Thomas precession in these coordinates. More precisely, a rotational motion is
non-isotropic and the axis of the rotation is a preferred direction which we can choose as $x_1$. Then we would need to
understand what do $\tl{x}_{2,3}$ represent. This should also help us to understand the precise link between the Snyder
basis and the $\kappa$-Minkowski basis in DSR theories.

A last remark to state that although $w$ behaves additively, it is not to be considered as a new notion of energy.
Indeed, we are still dealing at the level of the composition of velocities, i.e. the law describing the change of
reference frames, and not describing composite objects and scattering processes.

\medskip

Let us now discuss the relationship between gyro-groups and quantum groups.  Already using the coset structure deforms
the group multiplication and introduces a modified co-product which is reflected in a non-trivial law of composition of
speeds in our case. This deformation is encoded in the $h$ factors used to define the coset product. At this level,
there is two possible points of view.

On the one hand, the structures encountered here are supposed to $\kappa$-deformations  (of Poincar\'e groups), which
are properly understood as quantum groups. This is the usual point of view taken in the study of DSR theories.

On the other hand, from the present perspective, the $h$ factors are seen to  introduce a non-associativity and a
generalized (non-scalar) 3-cocycle. Through the relation $g_1*g_2=h_{12}\rhd(g_2*g_1)$, we see that the $R$-matrix,
identified as the operator $h_{12}$ acting on the coset, depends on the coset elements $g_1,g_2$, and thus there
doesn't seem to be a fixed quasi-triangular structure, but a dynamical one. There does exist a similar structure of
quasi-Hopf-algebras, where the structures are deformed by a non-trivial 3-cocycle which renders the algebra
non-associative. This hints towards a possible link with the 2-category formalism, which will be investigated in future
work. Nevertheless, we have seen that, in our case of $\SO(3,1)/\SO(3)$, there exists a basis (or equivalently a
parametrization) for which the coset product becomes associative. Thus it seems that the 3-cocycle should be trivial,
and then the gyro-group would be a quantum group. Based on this observation, we conjecture that gyro-groups should
generally admit an associative basis and thus a quantum group structure, unlike generic quasi-groups.

\subsection{Relative non-commutative geometry and the Notion of Observer}

The non-commutative structure of Special Relativity leads us towards the notion of {\it Relative Non-Commutative Geometry} \cite{relatncg}. First of all it is rather expected to consider Special Relativity as a theory of a non-commutative space. More precisely, the key point to go from Galilean to Lorentzian is the notion of simultaneity. From global it becomes local and relative to the observer. In our framework, the non-commutativity encodes this change and thus encodes the observer dependence.

The main issue is then to define exactly what is meant by an "observer". A priori, its definition at least requires its position and its speed. However, in our non-commutative framework, we face the same issue that in a quantum context and we also need to know the set of measurements that the observer can do. More exactly, we need to specify what the observer calls coordinates, which operators correspond to what the observer labels as energy or momentum. The choice of coordinates is about choosing a basis of the algebra and thus chooses a particular deformation. From that point of view, we might say that we decided to work out the theory as seen from the typical inertial observer. The theory as seen by an observer in an accelerated motion will certainly need to be described in different terms. Indeed, we know that the 3d frame attached to such an observer will get rotated along its trajectory due to the Thomas precession $h$ factors, so that the notion of coordinates will be defined differently.

A more direct proof of the observer dependence of the non-commutative structure is to see that the $h$ factor associated to a particular experiment actually depends on the observer. More precisely, let us consider two (pure) boosts $g_1$ and $g_2$. Then the Thomas precession factor is defined through $g_1g_2=hg_{12}$ where $g_{12}$ specifies the composed velocity. Considering another observer, which will measure the speed $\tl{g}_1$ instead of $g_1$. Then the composition of velocity will read $\tl{g}_1g_2=\tl{h}\tl{g}_{12}$ with $\tl{h}\ne h$. Another point, maybe more intuitive from the physical point of view, is a consequence of the non-commutativity of the space coordinates. Indeed as $[x_i,x_j]\ne 0$, we can not localize space points  exactly, and we have to write coherent states describing a semi-classical notion of space points. Then we expect two effects. The first is that the uncertainty and the shape of these states will changed with the velocity of the observer, or in other words, coherent states for one observer will not be coherent states for another one \cite{discrete}. This can be seen as related to the non-trivial action of the boosts in Special relativity, for example inducing length contraction. A second effect is that it is likely that the spread of the coherent states will depend of the distance of the space point to the origin, so that the location of the observer is relevant \cite{discrete}.

\medskip

These considerations lead to a slight change of perspective with respect to the usual approach to non-commutative geometry. Indeed usually one is working at the algebra level and deforming it. In this sense, one is working globally, with the full space at once. Here for the sake of interpretation one as to introduce the concept of observer and so a relative notion of localization. On the other hand one can change of observer and physics shouldn't be dependent on the choice of observer. This relativity principle is what we implemented when requiring that  the conservation laws be identical in every reference frame.

This notion of relative non-commutativity which appears already at the Special Relativity level, is more generally  useful when dealing with Quantum Gravity. Already, a first step towards a more complex theory is given by Deformed Special Relativity and, as it is shown in \cite{dsrgl}, in this extended context the definition of the observer or equivalently or a reference frame requires to specify its mass. More generally, when considering Quantum Gravity from a practical point of view, one needs to deal with partial observables and their evolution, which are obtained through partial gauge fixing of the full theory \cite{partialobs}. Indeed full/true (invariant) observables given directly as solution of all the constraints (among which the Hamiltonian constraint) are mostly unphysical due to their global nature. We need to localize them. This is achieved by introducing boundaries, and then through partial gauge fixing which specifies the physical meaning that the observer gives to the mathematical entities (like what he calls time and what he calls space coordinates). From this perspective, one needs therefore to introduce a notion of non-commutative geometry which depends on the observer. This reformulation of Special Relativity provides a simple example and further examples will be worked out in \cite{relatncg}.

\section{Lessons for Doubly Special Relativity}

Deformed Special Relativity arises when one considers Special Relativity and tries to implement an extra universal
bound  which arises from some quantum gravity effect. As we mentioned in the introduction, when introducing a new
universal maximal quantity which was initially not reachable, we can either change the symmetries or deform the current
symmetry. DSR does the latter. Interestingly, we can trace its first apparition back to Snyder's seminal work in 47 and
it got then rediscovered in the late 90's. DSR can be seen as making the  momentum space curved, which  is analogous to
the construction we introduced for the speed space. There are however a couple of problems that have arisen and met
controversy in the community. Following the SR case we propose some solutions for them. We however don't go in the
complete details, which we shall develop in a forthcoming article \cite{dsrgl}.

\subsection{DSR in short}
Let us recall quickly the DSR setting. One considers the Poincar\'e group, $\ppp= SO(3,1)\ltimes \R^4$. By construction
we have that $\ppp/SO(3,1)\sim \R^4$ is the translational  part, and  is identified with the momentum space. There are
two main ways to implement the maximal quantity (which  can be a  mass, or an energy or a trivector momentum):
geometric or algebraic. Ultimately the two approaches coincide, as shown by Kowalski-Glikman \cite{kg:lecture}.

The algebraic approach consists in keeping the Lorentz part untouched 
\begin{equation}\label{lorentz}
\begin{array}{rcl}
[M_i, M_j]&=& i\epsilon_{ijk}M_k, \;[N_i, N_j]= -i\epsilon_{ijk}M_k, \; [M_i, N_j]= i\epsilon_{ijk}N_k, \\
{[}M_i, p_j{]}&=& i\epsilon_{ijk}p_k, \; [M_i, p_0]= 0,
\end{array}
\end{equation}
while deforming the action of the boosts on the momentum
\begin{equation}\label{modlorentz}
\begin{array}{rcl}
[N_i, p_j]&=&A\delta_{ij}+B p_ip_j + C \epsilon_{ijk}p_k,   \\
{[}N_i, p_0{]}&=& D p_i,
\end{array}
\end{equation}
where $A, B, C, D$ are functions of $p_0, p_i^2, \kappa$.  We would like that the deformed Poincar\'e group becomes the
usual Poincar\'e group in the continuum limit where $\kappa\rightarrow \infty$. This gives therefore some conditions on
these functions ($A, D \rightarrow 1$, $B, C\rightarrow 0$). We can moreover show that the function $C$ has to be zero
from the Jacobi identity. Also from this latter we  have the differential equation
\begin{equation}
\frac{\partial A}{\partial p_0}D+ 2\frac{\partial A}{\partial \overrightarrow{p}^2}(A+ \overrightarrow{p}^2B)-AB=1.
\end{equation}
Different solutions of this equation, with the limit conditions for $\kappa\rightarrow\infty$ give us different
deformations.

These latter give rise to different physical situations.  It is not clear which one is the "true physical" one or if
there are all equivalent under some new physical principle. We come back to this ambiguity in the following subsection.
In any case, what is important is that one can incorporate a new universal bound while keeping the Lorentz symmetry
fine.

  The geometrical approach  consists in replacing the flat momentum
space by a curved space (a (anti) De Sitter space), as did Snyder so that $\R^4\rightarrow SO(4, 1)/SO(3,1)$ (resp.
$SO(3, 2)/SO(3,1)$ ). This de Sitter space is the new space of momenta. The four generators left (called the de Sitter
boosts) in the coset are identified with the momenta. We can embed this de Sitter space in the 5d Minkowski, using the
equation $\kappa^2= -\pi_0 ^2+\pi_1^2+\pi_2^2+\pi_3^2+\pi_4 ^2$, where the $\pi_A$ are the 5d Minkowski coordinates,
and $\kappa$ is a constant with dimension of mass. The Lorentz part of SO(4,1) is acting as usual on $\pi_A$,
\begin{equation}
\begin{array}{rcl}
[M_i, \pi_j]&=& i\epsilon_{ijk}\pi_k,  \; [M_i, \pi_0]=[M_i, \pi_4]=0, \\
{[}N_i, \pi_j{]}&=& \delta_{ij}\pi_0,  \; [N_i, \pi_0]= i\pi_i,  \; [N_i, \pi_4]=0.
\end{array}
\end{equation}
This is essentially the Snyder's approach and later Kowalski-Glikman found out that it was equivalent to the algebraic
approach \cite{kg:lecture}.

Indeed, when restricting the $\pi$'s to the homogenous space $SO(4, 1)/SO(3,1)$, in a particular coordinates system,
e.g. $\pi_0= \pi_0(p_0, \overrightarrow{p}), \pi_i = p_i \pi(\eta_0(p_0, \overrightarrow{p})), \pi_4 = \sqrt{\kappa-
\sum_{i=0}^{3}\pi_i^2 }$,  one  recovers the  commutation relations (\ref{lorentz}, \ref{modlorentz}), with the
functions $A,B,D$  expressed in terms of $\pi_0, \pi$. In this sense the two approaches are equivalent.

This is DSR in a  seedshell and now we sketch the problems associated to its physical interpretation and also their
solutions, in the light of the presentation of SR we just made.

\subsection{One or many deformations are physical?}
There are many possible deformations as we can see from (\ref{modlorentz}), and their physical significance is not
clear.

  From the algebraic point of view it seems that each deformation is physically distinguished. From the geometrical point of
view it  seems that the choice of coordinates (and so deformation) on the manifold is not physically relevant. The two
approaches seem therefore  to be contradictory.

The SR construction went into the same trouble but as we know pretty well how SR is working we were able to solve the
problem. One needs first to define the relevant physical quantities and then from there the coordinates systems are all
equivalent as expected. Basically it means that once the map $f:\R^3\arr H$, where $H$ is the hyperboloid for SR or
$g:\R^4\arr dS_{\kappa}$ in DSR has been defined then the physics is independent of the choice of coordinates. This map
$g$ should be determined, calibrated,  with the help of different experiments. They can consist in checking new
dispersion relations as mostly advocated, or can be of some new type, like measuring some new precession, a Lorentz
precession, alter-ego of the Thomas precession.

It is interesting to note that the interpretation of the  choice of the deformation can be refined. Indeed this choice corresponds to the choice of an observer. In our case we consider an {\it inertial observer}, which corresponds to the choice of the Snyder coordinates. A different observer, e.g. accelerated, would likely single out another deformation. This then prompts  to the mind a generalization of the equivalence principle: in the context of quantum gravity, one should expect that  all the deformations carry equivalent physics, but as seen by different observers. From this perspective, it is worth pushing further the study of the non-commutative formulation of SR and see how GR can arise in this algebraic deformation context. This would help to understand this new potential generalization  of the equivalence principle. We leave this point for further investigations.

\subsection{Non-associativity is physical or not?}

From the SR case, we draw two important lessons concerning non associativity. The first is that even if people take
associativity for granted, SR is an example where there is a physical evidence of non-associativity. So it can be
physical! This non-associativity can be derived mathematically  from the coset structure of the hyperboloid, and is
seen physically as the Thomas precession. One can do the same for the DSR case, as there is once again a coset
structure. The Thomas precession came up as the product of two boosts is a rotation times a boost. In DSR, we have that
the product of two de Sitter boosts  $T(p_i)$ is a Lorentz transformation $L$ times a de Sitter boost.
\begin{equation}\label{coset2}
T(p_1).T(p_2)= L(p_1, p_2)T (p_1\oplus p_2).
\end{equation}
One should find some experimental setting to measure it, and hopefully it could be measurable at scale much smaller
than the chosen maximum scale, just as the Thomas precession is seen at speeds which are not relativist (e.g. electron
around a nucleus). In this sense, the  lesson is that non-associativity is physical and even the physical evidence of a
new regime.

The second lesson that one should draw is that this non-associativity in SR is essentially linked to changes of reference frame. Indeed it is the notion of composition of velocities under change of reference frame that we initially modify, and not the notion of scattering. But then, of course, the notions of energy and momenta and their law of conservation under scattering do get modified in order to accommodate the principle of relativity in the deformed theory.
By analogy we propose the same interpretation for DSR: the deformed co-product should describe changes of reference frames. Then the bound introduced by DSR should be interpreted as a universal mass/energy scale and its compatibility with the co-product means that "if the mass/energy of one process is bounded by the (renormalized) Planck mass for one observer (in one reference frame) then it should be bounded in for any observers (in any reference frames)". It appears that one should deal in this context with the mass (and momentum) of the reference frame as well as its speed. We argue in a forthcoming paper \cite{dsrgl} why one should generically take into account the momentum of a reference (or its mass) and not only its speed. We are therefore introducing an extra refinement of the  relativity principle. DSR introduces a new object to describe the scattering, the {\it momentum pentavector}, which helps understanding the soccer ball problem as we shall see below. The law of scattering defined in terms of this penta-momentum will have then to be compatible with the relativity principle: the laws of conservation should stay the same under change of reference frames, and the reference frame should be described by their momentum.

The physical interpretation of the Lorentz precession is  very close to the one associated to the Thomas precession
(e.g. when the electron is turning around the nucleus). When changing of reference frame, there will be  in general  a
natural acceleration associated to this change and this acceleration  will generate a Lorentz transformation on the
initial reference frame, that will depend on the mass of the new reference frame.

It exists however some deformations which are associative (that is the Lorentz precession must be hidden) and where the
compositions of momenta is simpler: they correspond to the quantum group like deformations of the Poincar\'e group. It
is not clear to which physical situation they are associated to, i.e. to which observer. It is however clear that the
bicrossproduct basis or the $\kappa$-Minkowski basis are very close to the situation described in the section (III.A),
and a better understanding of this case will help a lot to understand the physical meaning of the $\kappa$-Minkowski
space.

Finally as a last comment, note that in these associative  basis there was a problem concerning the interpretation of
the coproduct as a scattering: it seemed that the whole universe was contributing to the scattering, due to the
non-cocommutativity of the coproduct. It is the so-called spectator problem \cite{kg:lecture}. However, we have seen that this co-product should be interpreted as describing the composition of momenta under changes of reference frames. Then we should first determine the scattering co-product, describing how to determine the total momentum of a composite object in terms of  the momenta of its components, and reconsider the issue of the spectator problem with respect to this new co-product \cite{dsrgl}.

This interpretation of the addition of momenta as a composition instead of a scattering is moreover corroborated by the
solution of the soccer ball problem, which we recall now.

\subsection{The soccer ball problem}
We are by construction imposing a constraint on the momenta. The quadrivector or some part of it cannot be bigger that
some universal scale. The addition of momenta is deformed so that it respects this bound. It happens however that in
the everyday experience, this bound is violated. For example if one one chooses to bound the energy and considers a
soccer ball, it is pretty obvious that this is violated. There seems to be something wrong right at the root of the
philosophy. The way out is to see that the deformed addition of  momenta corresponds not to a scattering but a
composition of momenta associated to different reference frames.

The scattering will be describe by some new objects associated to the new symmetry (the de Sitter group), just as the
scattering in the SR case is described not by the 3d Galilean momenta, but the 4d relativistic ones. In this sense, in
the DSR case one should consider the   "DSR momentum", which is now a pentavector, i.e. 5d vector.

In the Galilean setting, one considers the momentum trivector, and an extra component the energy, so in this sense
Galilean kinematics is described by a (3+1)d vector. In the relativistic case, we still have a 4d vector, but now space
and time are really on equal footing. One then might be inclined to say that we are adding an extra coordinate in the
DSR case by considering the pentavector. This is not the case. Indeed this fifth coordinate indicates the scale at
which we are working. Galilean physics is conformally invariant, just as Special Relativity. By implementing a
universal length, one is breaking this invariance and the one has to specify the scale at which one is working. This is
also a reason why physics of the DSR deformation is different than the SR one, as in the latter case, one is
implementing an universal ratio, a speed, which is then not breaking the conformal invariance.

Once defined, it is not difficult to see that the pentavector takes into account the maximum mass and that the trivial
scattering allows then to define a new maximum mass, solving naturally the soccer ball problem.

As we are defining a new momentum, we can expect to have some new physical features. Indeed, just as from the Galilean
perspective we had a new notion of energy and momentum which is emerging, in the DSR case we shall be able to define a
new notion of energy and momentum. All those new objects are described in the companion paper \cite{dsrgl}.

Magueijo and Smolin \cite{leejoao} had already a proposal for solving this soccer ball problem, but it was implying a
non-associative coproduct for the momenta and was moreover constructed by hand and so was falling out of the safe field
of quantum group deformations. Considering this effective momentum also allows to give a more mathematical
understanding of their trick.

\section*{Conclusion}
In the first sections, we have shown how Special Relativity can be constructed as a non commutative geometry from the
Galilean symmetries, instead of a commutative geometry arising from the Lorentz symmetries. We have explored the
different consequences of this construction in the light of the already existing Special Relativity. This allowed us to
settle some ambiguities. From the construction we made, it is clear that the DSR construction is  completely equivalent
to the one we made for SR, in particular, one gets  the same kind of ambiguities. It is then natural to proceed by
analogy and propose the same solutions as SR to settle down the DSR ambiguities. It allows in particular to tell us
that only one deformation is relevant to physics, and corresponds to the inertial observer, whereas the other are not
physical or concern different types of observers (accelerated...). It allows also to see that the non (co)associativity
is not a problem but a physical prediction of quantum gravity. It gave also some hints on the interpretation of the
modified addition of momenta. It shouldn't be interpreted as a scattering but more as composition of momenta under changes of reference frames. Indeed
reference frames should be now described by their speed but also by their mass. The scattering should be then described
by a new momentum associated to the new introduced symmetry (de Sitter group), just as scattering in relativistic
physics is not given by the scattering of Galilean momenta but the relativistic ones. We shall discuss in a companion
paper \cite{dsrgl} how this addresses both the soccer ball problem and the spectator problem.

We left also some open points for further investigations. The analogy between the DSR and SR cases  can be carried
further  out.  For example to understand the non isotropic coordinates (\ref{noniso}) and the Thomas precession should
help to understand the $\kappa$-Minkowski case. In fact, if one thinks of an accelerated observer there is a natural
preferred direction, and one can then conjecture that the bicrossproduct basis represents the case of an accelerated
observer. This statement is however under study.

Even at a further level, one can think about General Relativity viewed from the deformation point of view. Actually a
first step in this direction, i.e. considering a non flat space has been made in \cite{tagirov}. The same construction
applied for  the DSR case should be especially interesting, as it should represent an effective theory of Quantum
Gravity. At this stage one can argue that the correct mathematical tools to do this generalization would be the
Spectral Triples \cite{connes}, where one deals with the algebraic point of view. The deformation point of view  is
then easier to handle than the geometrical one. The Dirac operator encodes the metric of the manifold and it would
interesting to study the interplay of the deformation and this Dirac operator. This is especially appealing to us  as
it would renew the statement that non commutative geometry can be used to describe quantum gravity effects.

At a more mathematical level, we found that  quasigroup theory seems to be related to the study of  quantum groups.
This is  new to our knowledge.  In general quasigroups have drown less attention compared to the quantum groups. It is
then interesting to see that many of the features of the latter can be exported to the former for a better
understanding (e.g. representation theory...). We leave all these for further investigations.

As a final conclusion let us make a general comment on the introduction of different maximal quantities. We have seen
that we can deform the Galilean symmetries to mimic the Special relativity effects. On the other side, equivalently, to
go to Galilean to Lorentzian one is doing some complexification of the symmetries. Special Relativity in turn can be
also be deformed by introducing a minimal energy. One can therefore also introduce this maximal scale in the context of
the Deformed Galilean Relativity (DGR) and deform once again. One would obtain then a Doubly Deformed Galilean
Relativity (DDGR). We explored here the link between DGR and SR, but the link between DDGR and DSR would be also
interesting. Finally one could ask if instead of deforming one could have constructed some symmetries such that they
naturally encode the maximal energy and speed, while having a commutative underlying space-time. There exists in fact
one such candidate: one has to pseudo-complexify the Lorentz group\cite{BI}. This structure encodes then for example
the kinematics of the Born-Infeld action. We can then summarize this general picture through the following diagram.
$$
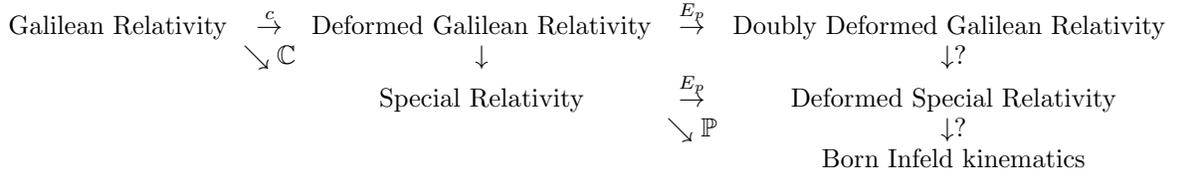
\begin{array}{lcccc}
\textrm{Galilean Relativity} & \stackrel{ c}{\rightarrow}& \textrm{Deformed Galilean Relativity}& \stackrel{E_p}{\rightarrow} &\textrm{Doubly Deformed Galilean Relativity }  \\
& \searrow \C & \downarrow  & & \downarrow ? \\
&& \textrm{Special Relativity} &\stackrel{E_p}{\rightarrow} &  \textrm{Deformed Special Relativity}  \\
&&& \searrow \P & \downarrow ?  \\ &&&& \textrm{Born Infeld kinematics}
\end{array}$$

\section*{Acknowledgement}
FG would like to thank Frederic Schuller for a discussion at Christmas dinner.
EL would like to thank Laurent Freidel for many discussions and explanations on the quantum group
structure of DSR.

\appendix

\section{Speed composition and Thomas Precession in  Special Relativity}
In this section, we describe how one gets the speed composition from product of boosts. Let be given two boosts $g_1$ and $g_2$. In the spinorial representation, their product read:
\bes
g_1g_2&=& \left(\cosh\frac{\eta_1}{2} Id+ \sinh\frac{\eta_1}{2} \overrightarrow{b_1}.
\overrightarrow{J}\right)
\left(\cosh\frac{\eta_2}{2} Id+ \sinh\frac{\eta_2}{2}
\overrightarrow{b_2}. \overrightarrow{J}\right)\nn\\
&=& \left(\cosh\frac{\eta_1}{2}\cosh\frac{\eta_2}{2}+ \sinh\frac{\eta_1}{2}\sinh\frac{\eta_2}{2}
\overrightarrow{b_1}.\overrightarrow{b_2}\right)\,Id \nn\\
&& + \left(\cosh\frac{\eta_2}{2}\sinh\frac{\eta_1}{2}\v{b_1}+ \cosh\frac{\eta_1}{2}\sinh\frac{\eta_2}{2}\v{b_2}+i\sinh\frac{\eta_1}{2}\sinh\frac{\eta_2}{2}(\v{b_1}\wedge\v{b_2})
\right).\v{J}
\ees
This product is then a element of $\SO(3,1)$, which can be factorized as the product of a rotation times a boost (Cartan decomposition). Noting the rotation $h= \cos\theta + i\sin\theta\v{r}.\v{J}$ and $g$ the total boost, we have:
\bes
hg&=& \left(\cos\theta Id+ i\sin\theta \overrightarrow{r}. \overrightarrow{J}\right)\,\left(\cosh\frac{\eta}{2} Id+ \sinh\frac{\eta}{2} \overrightarrow{b}. \overrightarrow{J}\right)\nn\\
&=&\left(\cos\frac{\theta}{2} \cosh \frac{\eta}{2} +i \sin\frac{\theta}{2}\sinh\frac{\eta}{2}
\overrightarrow{r}.\overrightarrow{b}\right)\, Id \nn\\
&& +\left(\cos\frac{\theta}{2}\sinh\frac{\eta}{2}\overrightarrow{b}+i
\sin\frac{\theta}{2} \cosh\frac{\eta}{2} \overrightarrow{r}-
\sin\frac{\theta}{2}\sinh\frac{\eta}{2}(\overrightarrow{r}\wedge\overrightarrow{b})\right).\overrightarrow{J}
\ees
Identifying the two previous expressions gives us a set of equations from which we deduce the composition of the speeds and the Thomas precession:
\bes
i \sin\frac{\theta}{2}\sinh \frac{\eta}{2} \overrightarrow{r}.\overrightarrow{b}&=& 0 \\
\cos\frac{\theta}{2}\cosh \frac{\eta}{2} &=& \cosh\frac{\eta_1}{2}\cosh\frac{\eta_2}{2}+ \sinh\frac{\eta_1}{2}\sinh\frac{\eta_2}{2} \overrightarrow{b_1}.\overrightarrow{b_2}\\
\cos\frac{\theta}{2}\sinh\frac{\eta}{2}\overrightarrow{b}-
\sin\frac{\theta}{2}\sinh\frac{\eta}{2}(\overrightarrow{r}\wedge\overrightarrow{b})&=& \cosh\frac{\eta_2}{2}\sinh\frac{\eta_1}{2}\overrightarrow{b_1}+\cosh\frac{\eta_1}{2}\sinh\frac{\eta_2}{2}\overrightarrow{b_2}\\
\sin\frac{\theta}{2} \cosh\frac{\eta}{2} \overrightarrow{r}&=&
\sinh\frac{\eta_1}{2}\sinh\frac{\eta_2}{2}(\overrightarrow{b_1}\wedge\overrightarrow{b_2}).
\ees
First, we have $\v{r}$ orthogonal to $\v{b}$. And dividing the last equation by the second one, we get the value of the rotation:
\begin{equation}
\tan\frac{\theta}{2} \overrightarrow{r}
=\frac{\tanh\frac{\eta_1}{2}\tanh\frac{\eta_2}{2}}{1+\tanh\frac{\eta_1}{2}\tanh\frac{\eta_2}{2}
\overrightarrow{b_1}.\overrightarrow{b_2}}(\overrightarrow{b_1}\wedge\overrightarrow{b_2}).
\end{equation}
Then dividing the third equation by the second, we get:
$$
\left(Id-\tan\f{\theta}{2}\,\v{r}\wedge
\right)\,
\tanh\frac{\eta}{2}\v{b}
=\frac{\tanh\frac{\eta_1}{2}\v{b_1}+\tanh\frac{\eta_2}{2}\v{b_2}}{1+\tanh\frac{\eta_1}{2}\tanh\frac{\eta_2}{2}
\overrightarrow{b_1}.\overrightarrow{b_2}}.
$$
Taking into account than $\v{r}$ is orthogonal to all $\v{b_1},\v{b_2},\v{b}$, we can invert the operator
$Id-\v{R}\wedge$ and obtain that:
\be
\tanh\frac{\eta}{2}\v{b}=
\cos^2\f{\theta}{2}\,
\left(1+\tan\f{\theta}{2}\,\v{r}\wedge\right)\,
\frac{\tanh\frac{\eta_1}{2}\v{b_1}+\tanh\frac{\eta_2}{2}\v{b_2}}{1+\tanh\frac{\eta_1}{2}\tanh\frac{\eta_2}{2}
\overrightarrow{b_1}.\overrightarrow{b_2}}.
\ee
A faster way to the value of $\eta$ is to directly apply the boost $g_2$ to the point $(\cosh\eta_1,\sinh\eta_1\v{b_1})$ on the hyperboloid. This way, we get:
\bes
\cosh\eta&\equiv& \f{1}{2}\textrm{Tr}(g_2^\dagger g_1^\dagger g_1g_2)
=\f{1}{2}\textrm{Tr}(g_1^2g_2^2) \nn\\
&=& \cosh\eta_1\cosh\eta_2+\sinh\eta_1\sinh\eta_2\v{b_1}.\v{b_2}.
\ees
Finally,  we translate all these formulas to velocities using the definition that $\v{v}=cf(\eta)\v{b}$.



\begin{thebibliography}{99}

\bibitem{DSR}
G Amelino-Camelia, {\it Are we at the dawn of quantum-gravity phenomenology?}, Lect.Notes Phys. 541 (2000) 1-49;


\bibitem{dsrcosmo}
G Amelino-Camelia, L Smolin, A Starodubtsev, {\it Quantum symmetry, the cosmological constant and Planck scale
phenomenology}, hep-th/0306134;

\bibitem{dsr3d}
L Freidel, J Kowalski-Glikman, L Smolin, {\it 2+1 gravity and Doubly Special Relativity}, hep-th/0307085;

\bibitem{qg-effect}
F Girelli, ER Livine, D Oriti, {\it Deformed Special Relativity as an effective flat limit of quantum gravity},
gr-qc/0406100;

\bibitem{deformedgalilee}
P Kosi\'nski, P Ma\'slanka, {\it Deformed Galilei symmetry}, math.QA/9811142;

\bibitem{snyder}
H. Snyder, {\it Quantized space-time}, Phys. Rev. {\bf 71} (1947) 38;

\bibitem{discrete}
ER Livine, D Oriti, {\it About Lorentz invariance in a discrete quantum setting}, J. High Energy Phys. {\bf 06} (2004)
050, gr-qc/0405085

\bibitem{thomas}
L Thomas,  Phil. Mag. 3, (1927), 1


\bibitem{ungar}
J Chen, A Ungar,
{\it From the group $SL(2,\C)$ to gyrogroups and gyrovector spaces and hyperbolic geometry}, http://www.math.ndsu.nodak.edu/faculty/ungar/publications.html; \\
A. Ungar, {\it Thomas rotation and the parametrization of the Lorentz transformation group}, Found. Phys. Lett.  1
57-89;

\bibitem{quasigroups}
JDH Smith, {\it Loops and quasigroups: Aspects of current work and prospects for the future},
Comment.Math.Univ.Caroliane 41, 2 (2000), 415-427

\bibitem{3cocycle}
AI Nesterov, {\it Three-cocycles, Nonassociative Gauge Transformations and Dirac's Monopole}, hep-th/0406073

\bibitem{relatncg}
F Girelli, ER Livine, {\it Relative non commutative geometry}, in preparation;


\bibitem{dsrgl} F Girelli, ER Livine, {\it DSR: about the soccer ball problem and a Lorentz precession}, in
preparation;

\bibitem{partialobs}
C Rovelli, {\it Partial Observables}, gr-qc/0110035, Phys.Rev. {\bf D65} (2002) 124013

\bibitem{kg:lecture} J Kowalski-Glikman, {\it Introduction  to Doubly Special Relativity}, hep-th/0405273;


\bibitem{leejoao}
J Magueijo, L Smolin, {\it Generalized Lorentz invariance with an invariant energy scale}, Phys.Rev. {\bf D67} (2003)
044017, gr-qc/0207085;

\bibitem{tagirov}
EA Tagirov, {\it Effective Space Quantization in Friedman-Robertson-Walker Models}, gr-qc/0312074

\bibitem{BI}
F Sch\"uller, {\it Born-Infeld kinematics}, Annals Phys. {\bf 299} (2002) 174-207, hep-th/0203079

\bibitem{connes} A Connes, {\it A Short Survey of Noncommutative Geometry}, J.Math.Phys. 41 (2000) 3832-3866, hep-th/0003006






\end{thebibliography}
\end{document}